\documentclass{pasj00}
\usepackage{times}
\usepackage{dcolumn}

\begin{document}
\SetRunningHead{H. Nakanishi and Y. Sofue}{3-D Distribution of Molecular Gas in
the Milky Way Galaxy}
\Received{2006/02/14}
\Accepted{2006/08/03}

\title{Three-Dimensional Distribution of the ISM in the Milky Way
Galaxy: II. The Molecular Gas Disk}
\author{Hiroyuki \textsc{Nakanishi}\altaffilmark{1} and Yoshiaki \textsc{Sofue}\altaffilmark{2}}%
\altaffiltext{1}{Nobeyama Radio Observatory, Minamimaki, Minamisaku, Nagano 384-1305}
\altaffiltext{2}{Institute of Astronomy, The University of Tokyo, 2-21-1 Osawa, Mitaka,Tokyo 181-0015}
\email{hnakanis@nro.nao.ac.jp}
\maketitle

\begin{abstract}
We created a three-dimensional distribution map of molecular gas throughout the Milky Way Galaxy using the latest $^{12}$CO($J=$1--0) survey data cube and rotation curve based on the kinematic distance. 
The radial distribution of the molecular gas shows a central peak and a second peak around 0.5 $R_0$. The thickness of the molecular disk slightly increases from 48 pc to 160 pc with the galactocentric distance within a radius range of 0--11 kpc. 
We were able to trace the Outer, the Perseus, the Sagittarius-Carina, the Scutum-Crux, and the Norma arms as logarithmic spiral arms with pitch angles of $11\arcdeg - 15\arcdeg$. Considering that the pitch angles of the spiral arms are within this range, the Norma and the Outer arms {seem to} be identified as the same spiral arm. 
We could also trace a midplane displacement, whose amplitude is nearly constant inside a 10 kpc radius and increases beyond this radius. The ridges of midplane displacement form leading spiral arms. 
\end{abstract}

\section{Introduction}
The global three-dimensional (3D) structure of the Milky Way Galaxy is one of the essential informations to study the Galactic objects in detail. 
In Paper I \citep{nak03}, we have analyzed the latest 21-cm line survey data, and presented the 3D distribution map of H {\sc i} gas in the Milky Way Galaxy. 
In this paper, we deal with the molecular gas distribution as Part II of this series. 

A survey in the $^{12}$CO($J=$1--0) line is a good tool for tracing the interstellar matter (ISM) in the inner disk of the Milky Way Galaxy (inside the solar orbit), while  the H {\sc i} data is advantageous for study of ISM in the outer disk (outside the solar orbit). 
One of the most famous CO line surveys is the Columbia CO survey, whose earlier survey was carried out with the $1.2$-m telescopes in New York City, USA, and Cerro Tololo, Chile \citep{dam87,bro88,bro89}. This survey covered the entire Milky Way with a grid spacing of $0\fdg5$. Another higher resolution survey is the Massachusetts--Stony Brook Galactic plane CO survey, which was made using the 14-m telescope of the Five College Radio Astronomy Observatory (FCRAO) \citep{san86}. The covered ranges of the galactic longitude and latitude were from $l=8\arcdeg$ to $ 90\arcdeg$ and from $b=-1\fdg05$ to $+1\fdg05$ in the first quadrant, respectively. The grid spacing was $3\arcmin \times 3\arcmin$ or $6\arcmin \times 6\arcmin$. 
These large CO surveys were ideal for exploring into the molecular gas distribution in the Milky Way Galaxy. There have been several studies on the Galactic structure using these data. 

\citet{dam86} and \citet{mye86} showed face-on views of the first Galactic quadrant examining positions of the largest individual molecular complexes identified in the Columbia CO survey of the first Galactic quadrant \citep{coh86}. They found clear evidence for the Sagittarius spiral arm. \citet{coh85}, \citet{gra87}, and \citet{gra88} showed distribution of giant molecular clouds in the fourth quadrant and they clearly delineated the Carina arm. 
\citet{cle88} presented the molecular gas distribution in the first quadrant of the inner disk of the Milky Way Galaxy using the Massachusetts--Stoneybrook CO survey, found a ring-like structure at the radius of $0.5 R_0$ ($R_0$: the Galactocentric distance of the Sun), and traced the Sagittarius and Perseus arms. The distance of the molecular gas was determined by the kinematic distance using a rotation curve. The near--far problem in the inner Galaxy was {overcome} by introducing a vertical model of the molecular gas distribution. \citet{sol89} showed a face-on view of the Milky Way Galaxy examining individual giant molecular clouds identified in the Massachusetts-Stoneybrook CO survey.

\citet{hun97} presented the molecular gas distributions in the entire Galactic disk of the Milky Way Galaxy using the Columbia survey \citep{dam87} for modeling the diffuse gamma-ray emission. The distance of the molecular gas was determined by the kinematic distance. Although there is a near--far problem in the inner disk, they {\bf divided} gas density equally at near and far points for $b=0\arcdeg$. 

\citet{saw04} determined the molecular gas distribution in the Galactic center by considering the absorption of the OH line to determine the line-of-sight distance of the molecular gas, using the CO data cube of \citet{bit97}. They found that molecular gas is elongated in the line-of-sight direction against the Sun, and that this feature is associated with a central galactic bar.

Thus, there are several studies on the molecular gas distribution. However, there has been no three-dimensional {gas density} map of the whole Galaxy. 
\citet{dam01} {published a new compilation of } Columbia $^{12}$CO($J=$1--0) survey of the whole of the Milky Way Galaxy. The sampling and the sensitivity of this survey were much improved over that of the old Columbia survey. This is a great tool for investigating the molecular gas distribution throughout the entire Milky Way Galaxy. 
In this paper, we show a molecular gas distribution in the Milky Way Galaxy using the Columbia $^{12}$CO($J=$1--0) survey data and the latest rotation curve.

\section{DATA}
We used the latest {compilation of } $^{12}$CO($J=$1--0) survey data \citep{dam01}, which were obtained with the 1.2-m telescopes in New York City, USA, and Cerro Tololo, Chile. The half-power beam width (HPBW) was $8\farcm4$, which corresponded to 2.4 pc at the heliocentric distance of 1 kpc. 
The data consisted of $^{12}$CO($J=$1--0) line spectra over the entire {Milky Way. We used data of surveys named 2, 8, 18, 31, 33, and 36. Grid spacing of these data was $0\fdg125$ -- $0\fdg25$. }
The total radial velocity ($V_{\rm r}$) coverage was 332 km s$^{-1}$ and velocity resolution was $1.3$ km s$^{-1}$. The r.m.s. noise level per channel was 0.12 -- 0.43 K. We used spectra within a Galactic latitude range of $|b| \le 1.5\arcdeg$.

The rotation curve, which we used to transform the radial velocity of the CO gas into the distance, is the same as we used in Paper I. The inner and outer rotation curves are taken from \citet{cle85} and \citet{deh98}, respectively. The Galactic constants $R_0 = 8.0$ kpc (the Galactocentric distance of the Sun) and $V_0 = 217 $ km s$^{-1}$ (the solar rotational velocity) are adopted.

\section{METHOD}
\subsection{Binding the Data}
\subsubsection{Inner Galaxy}
For the inner Galaxy, we averaged the CO data over five channels ($\ge 6.5$ km s$^{-1}$ velocity width) for each spectrum in order to eliminate the cloud-cloud motion of 3 -- 3.9 km s$^{-1}$\citep{cle85,alv90} and to obtain a higher signal-to-noise ratio (S/N ratio). 

The molecular gas is distributed in a more clumpy form than the H {\sc i} gas. Since we intend to develop our understanding of the global structure of the Milky Way Galaxy, we must smooth the CO emission on a larger scale in order to eliminate such a clumpy features and to model the vertical structure of the molecular gas disk. Since the typical scale of the random motion of the molecular clouds is $0.015R_0$ kpc \citep{cle85}, we averaged the data over a 240 pc width when the line-of-sight length corresponding to 6.5 km s$^{-1}$ velocity width was smaller than 240 pc. Thus the CO data were averaged over $\ge 6.5$ km s$^{-1}$ velocity width and $\ge240$ pc width. 

In addition, we averaged the data in the longitudinal direction to eliminate the clumpy features of CO clouds and to obtain a much higher S/N ratio. For the first ($0\arcdeg \le l \le 90\arcdeg$) and fourth ($270\arcdeg \le l \le 360\arcdeg$) quadrants, we bound the data, so that the longitudinal width at the tangential point is equal to 240 pc-width.

Thus the CO data were averaged over $\ge 240$ pc in the line-of-sight direction and longitudinal direction except for the inner region of the tangential radius, where the data were averaged over $\ge 240$ pc width only in the line-of-sight direction. The data were not averaged in the latitudinal direction at all.

\subsubsection{Outer Galaxy}
For the outer Galaxy, we averaged the data in the longitudinal direction so that longitudinal separation becomes $0\fdg5$, which is nearly the same as the H {\sc i} data. 
Moreover, we averaged the CO data over two channels ($\ge 2.6$ km s$^{-1}$ velocity width), which is also the same as the H {\sc i} data.

Figure \ref{cal-point} shows the points where the molecular gas densities were calculated. 

\subsection{Molecular Gas Distribution in the Outer Galaxy}
In this paper, we choose a Cartesian coordinate whose $x$--axis coincides with the line crossing the Sun and the Galactic center and whose origin coincides with the Galactic center. The $z$-axis is chosen so that it is parallel to the rotational axis. The Sun is located at $(-8 \mbox{ kpc}, 0, 0)$. We also choose a cylindrical coordinate $(R, \theta, z)$ so that the angle $\theta=180\arcdeg$ coincides with the direction toward the Sun and the angle $\theta=90\arcdeg$ is parallel to $l=90\arcdeg$. 

The distance of the molecular gas was determined by kinematic distance in this paper, the same way as in Paper I. 
The distance of molecular gas in the outer disk $(R > R_0)$ is uniquely determined, and then we obtained the molecular gas distribution in the outer disk using the equation
\begin{equation}
n_{{\rm H}_2} = X T_{\rm b} {\Delta V_{\rm r} \over \Delta r}
\end{equation}
where $n_{{\rm H}_2}$ denotes the volume number density of the molecular hydrogen H$_2$, $X$ the CO-to-H$_2$ conversion factor, and $r$ the heliocentric distance.  
We adopted two kinds of conversion factor; (1) a constant conversion factor of $X=1.8 \times 10^{20}$ (H$_2$ cm$^{-2}$ K$^{-1}$ km$^{-1}$ s) \citep{dam01} and (2) a varying conversion factor with the Galactocentric distance \citep{ari96}. 
\citet{ari96} calculated $X$ using the data in \citet{sol87}. 

Since the conversion factor can be approximately expressed by exponential function of the Galactocentric radius, we fitted an exponential function to the resultant radial profile and obtained the following equation,
\begin{equation}
\label{conv}
X \mbox{[H$_2$ cm$^{-2}$ K$^{-1}$ km$^{-1}$ s]} = 1.4 \exp{(R/11 {\rm kpc})}. 
\end{equation}

\subsection{Molecular Gas Distribution in the Inner Galaxy}
There is, on the other hand, a near-far ambiguity in the inner disk $(R > R_0)$; i.e. there are two heliocentric distances satisfying the same radial velocity. Therefore, we could not determine the H$_2$ distribution in the inner Galaxy in the same way as in the outer Galaxy. 
{In order to divide molecular distributions at near and far points, we introduced} the vertical structure of the Galactic disk. Following \citet{spi42}, we assumed that the vertical distribution of the molecular gas is expressed with 
{
\begin{eqnarray}
\label{spi_eq}
n(z) = n_0 {\rm sech}^2(\xi), \\
\xi = \log{(1 + \sqrt{2})}{z - z_0 \over z_{1/2}}, 
\end{eqnarray}
}
where $z$ and $z_{1/2}$ denote the height from the Galactic plane ($b=0\arcdeg$) and the scale-height, respectively. At $z=z_{1/2}$ the density, $n$, gets a half value of the maximum at the midplane. The subscript $0$ means values at the midplane. The scale-height was determined at the tangential points in the inner Galaxy, which is described in the next subsection. 

Assuming that the vertical distribution of the H$_2$ gas follows equation (\ref{spi_eq}), the H$_2$ column density, $N_{{\rm H}_2}(b)$, is represented as a function of the Galactic latitude $b$ using four parameters: $n_{{\rm H}_2{1}}$, $n_{{\rm H}_2{2}}$, $z_{0_1}$, and $z_{0_2}$. That is, 
\begin{equation}
N_{{\rm H}_2}(b) = n_{{\rm H}_2{1}} {\rm sech}^2(\xi_1) {\Delta r_1
\over \cos{b}} + n_{{\rm H}_2{2}} {\rm sech}^2(\xi_2) 
{\Delta r_2 \over \cos{b}}, 
\end{equation}
where
\begin{equation}
\xi_1 = \log{(1+\sqrt{2})}{r_1 \tan{b} - z_{0_1}\over z_{1/2}}, 
\end{equation}

\begin{equation}
\xi_2 = \log{(1+\sqrt{2})}{r_2 \tan{b} - z_{0_2}\over z_{1/2}}. 
\end{equation}

For each point {shown in figure \ref{cal-point}}, we sought a parameter set consisting of $n_{{\rm H}_2{1}}$, $n_{{\rm H}_2{2}}$, $z_{0_1}$, and $z_{0_2}$, which gives the smallest residual in subtracting the model distribution from the observation. 
The fitting resolution of the four parameters is 0.01 H$_2$ cm$^{-3}$ for $n_{{\rm H}_2{1}}$ and $n_{{\rm H}_2{2}}$, and 1 pc for $z_{0_1}$ and $z_{0_2}$, respectively. We restricted $z_{0_1}$ and $z_{0_2}$ within a range of $-250$ pc to $+250$ pc. 
{The grid spacing is $0\fdg125$, which corresponds $<17$ pc at tangential points, at most points in the inner Galaxy. Since there are enough points (more than three points) to resolve near vertical distribution, it is possible to divide the total CO emission into near and far emissions. }
We thus obtained parameter sets of $n_{{\rm H}_2{1}}$, $n_{{\rm H}_2{2}}$, $z_{0_1}$, and $z_{0_2}$ for discrete points in the inner Galaxy. 

\begin{figure*}
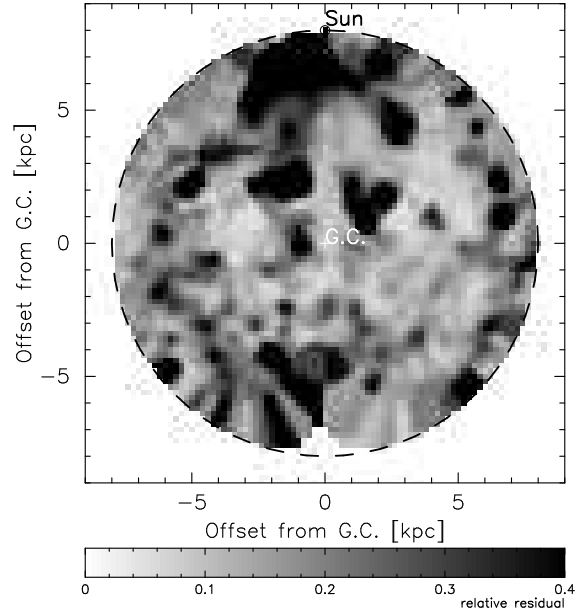

  \begin{center}
    \FigureFile(75mm,75mm){hnakanis-f4.ps}
  \end{center}
\caption{Residual map. We here define the 'relative residual' to be a fraction of standard deviation of residuals between fitted model curves and observation data relative to the peaks of the observation data. \label{residualmap}}
\end{figure*}

{To show how well the fitting method works, we present a residual map in figure \ref{residualmap}. In order to make this map, we first measured standard deviation of residuals between a fitted model curve and observed data. And then, we calculated relative residuals, which is a fraction of the standard deviation relative to the peak of the observational data. This residual map shows that the relative residual is less than 20 -- 30 \% in most of the region, and that the fitting method does not work well near the Sun. }

\subsection{Scale-height}
Tangential points which satisfy $R=R_0 \sin{l}$ present no near-far problem in the inner Galaxy. Hence, the vertical structure at the tangential points can be determined uniquely. 

\begin{figure*}
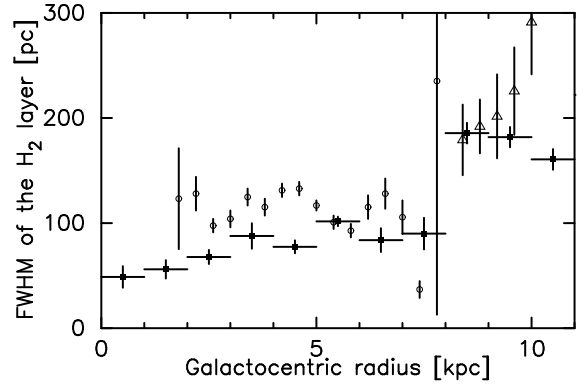

  \begin{center}
    \FigureFile(75mm,150mm){hnakanis-f1.ps}
  \end{center}
  \caption{FWHM of the H$_2$ gas layer versus the Galactocentric radius. Filled squares, open circles, and open triangles denote the data from this study, \citet{bro88} and \citet{gra87}, respectively. \label{sh-co}}
\end{figure*}

The gas of velocity range $|V_{\rm r, max}- \sigma | < |V_{\rm r}| < \infty$ is located at the tangential points, where $V_{\rm r}$, $V_{r, max}$, and $\sigma$ are radial velocity, maximum radial velocity, and random velocity, respectively. The random velocity was taken to be 5 km s$^{-1}$ \citep{cle85}. We thus obtained vertical distributions, and fitted a model distribution $\rho (z) = \rho_0 {\rm sech}^2(\xi)$ to determine a scale-height of $z_{1/2}$. Figure \ref{sh-co} shows a full-width half maximum (FWHM) ($2z_{1/2}$) plotted versus the Galactocentric radius, which is a mean value in a 1 kpc bin. {For a comparison, figure \ref{sh-co} shows also FWHM of molecular disk calculated by \citet{bro88} and \citet{gra87}, who use the same data that we use. We adopt the Galactic constant $R_0$ of 8 kpc although  \citet{bro88} and \citet{gra87} use $R_0$ of 10 kpc. Therefore, the radius and FWHM are reduced by factor of $8/10$. Filled squares, open circles, and open triangles denote the data from this study, \citet{bro88} and \citet{gra87}, respectively. Our result is almost consistent with \citet{bro88} and \citet{gra87}.}

\begin{figure*}
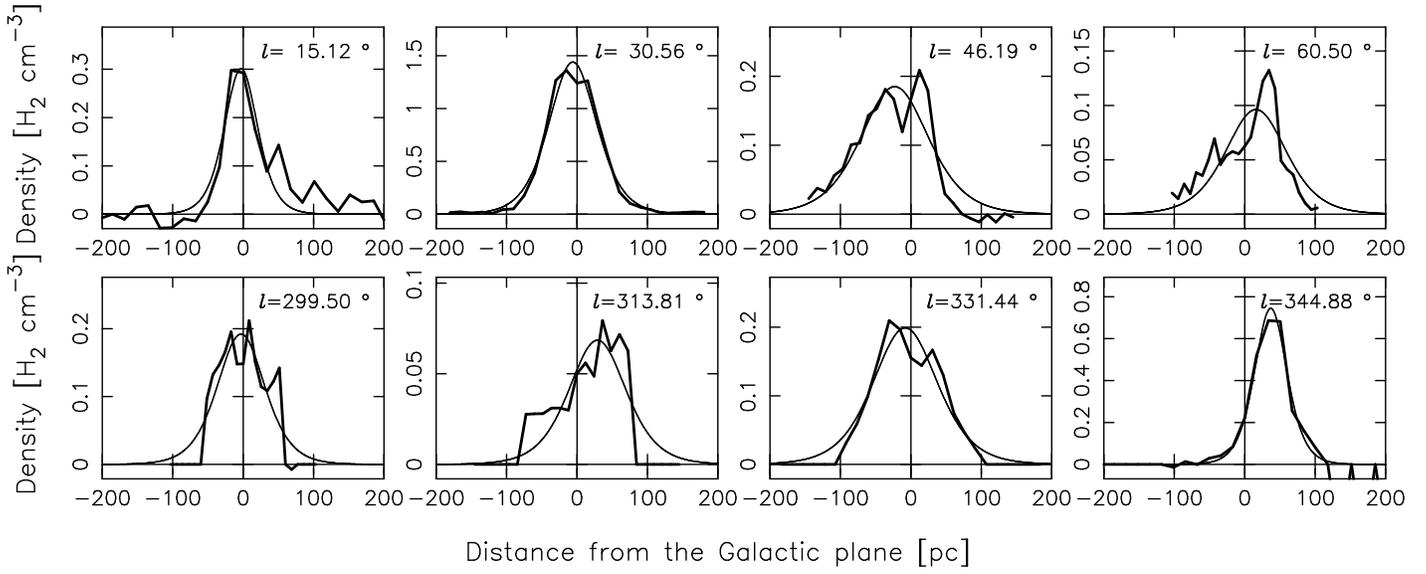

  \begin{center}
    \rotatebox{-90}{\FigureFile(75mm,75mm){hnakanis-f3.ps}}
  \end{center}
\caption{Vertical distributions of the molecular gas. Horizontal and vertical axes denote height from the Galactic plane ($b=0\arcdeg$) and H$_2$ density, respectively. Thick and thin lines indicate observation data and model curves, respectively.  \label{obs-model-tang}}
\end{figure*}

{Figure \ref{obs-model-tang} shows comparisons of a model and observations. A vertical distributions of molecular gas can be approximately reproduced by equation (\ref{spi_eq}).}

\vskip 3mm
\centerline{--- Figure \ref{sh-co} ---}
\vskip 3mm

\subsection{Construction of Three-Dimensional Map}
\begin{figure*}
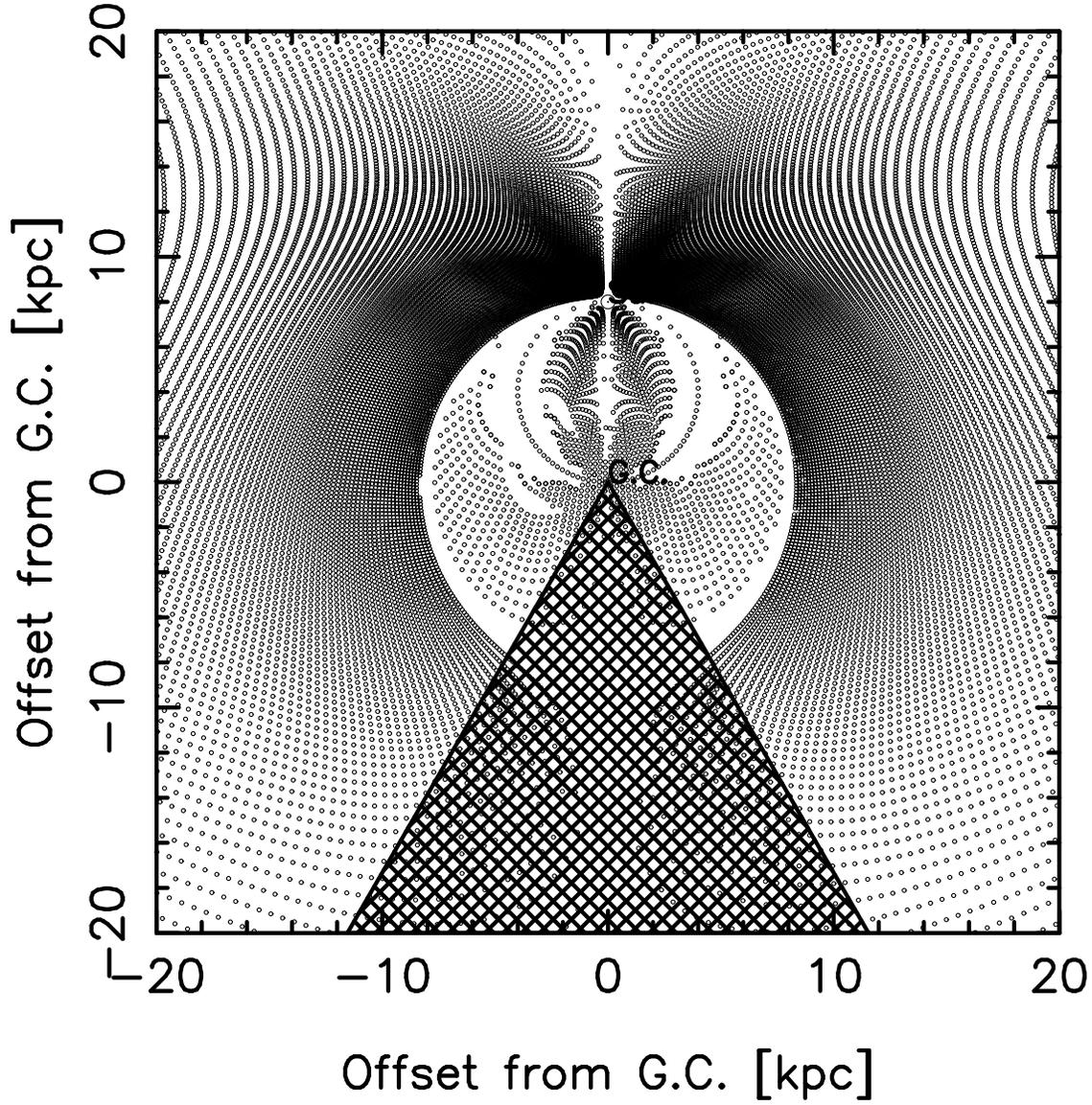

  \begin{center}
    \FigureFile(150mm,150mm){hnakanis-f2.ps}
  \end{center}
\caption{Grid points where gas densities were calculated. The hashed area denotes a region where there are too few points to make a density map.  \label{cal-point}}
\end{figure*}

We thus obtained parameter sets of the molecular distribution at each point of figure \ref{cal-point}.

First, the parameter sets obtained for the discrete points in the inner Galaxy were averaged with a Gaussian function whose FWHM is 0.48 kpc, in order to eliminate {data with bad fitting results.}
Then we calculated the H$_2$ distribution in the inner Galaxy using the averaged parameter sets. 
The H$_2$ gas densities obtained at the discrete points in the outer Galaxy were transformed into continuous distribution by interpolating with the same Gaussian function. Averaging length for the $z$-direction was fixed at 35 pc, which corresponds to the grid spacing of $0\fdg25$ at the heliocentric distance of 16 kpc. 
Finally, we combined the H$_2$ distributions in the inner and outer Galaxy to obtain a CO distribution cube. In this paper, the grid spacing of the cube is fixed at 200 pc for the $x$ and $y$ directions and at 20 pc for the $z$ directions, and the size of the resultant cube is 40 kpc $\times$ 40 kpc $\times$ 2 kpc in $x$, $y$, and $z$ direction, respectively.  

Since the number of points where the molecular gas densities were calculated (figure \ref{cal-point}) in the azimuthal range $|\theta|<30\arcdeg$ is too few to make a map, we do not show this region. 

\vskip 3mm
\centerline{--- Figure \ref{cal-point} ---}
\vskip 3mm

\section{Results}

\subsection{A Face-on Map}
\begin{figure*}
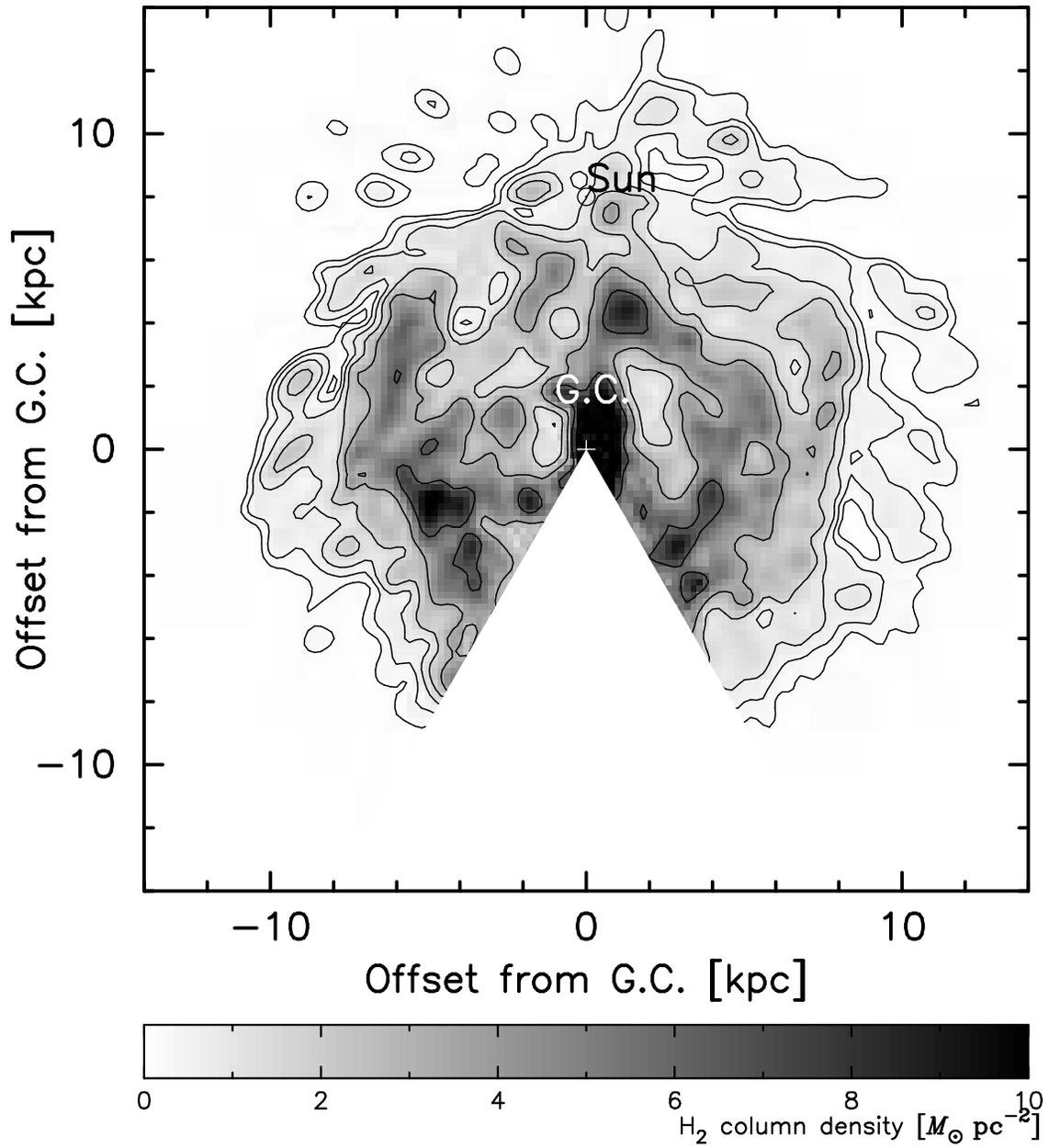

  \begin{center}
    \FigureFile(150mm,150mm){hnakanis-f5.ps}
   \end{center}
\caption{Face-on molecular hydrogen map adopting the constant conversion factor of $X=1.8 \times 10^{20}$ \citep{dam01}. The contour levels are 0.2, 0.4, 0.8, 1.6, 3.2, 6.4, and 12.8 $M_\odot$ pc$^{-2}$. \label{H2map1}}
\end{figure*}

Figure \ref{H2map1} shows a face-on map of the molecular H$_2$ gas in the Milky Way Galaxy. 

\begin{figure*}
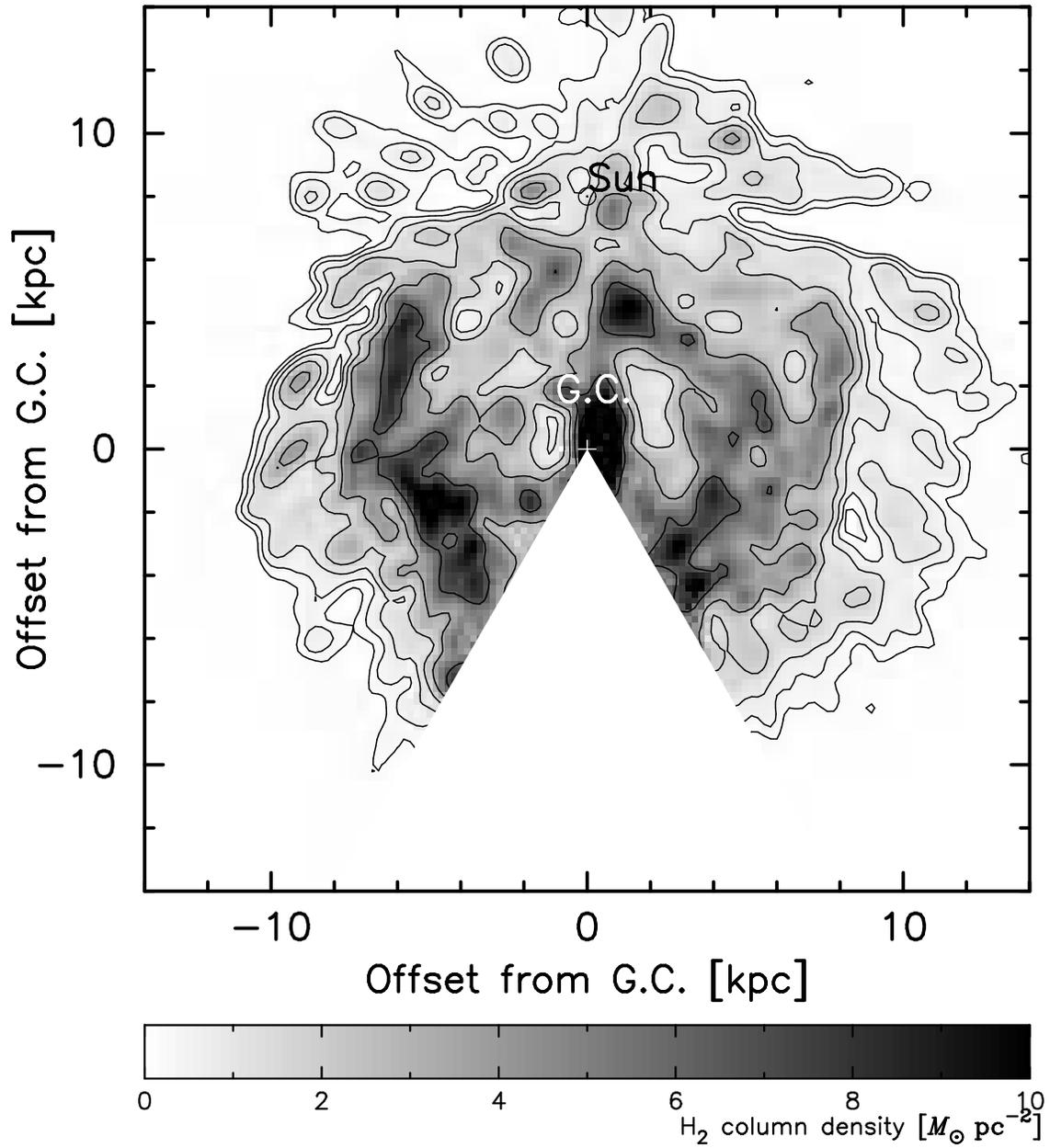

  \begin{center}
    \FigureFile(150mm,150mm){hnakanis-f6.ps}
   \end{center}
\caption{Face-on molecular hydrogen map adopting the conversion factor $X$ taken from \citet{ari96}, where $X$ increases exponentially with the galactocentric distance. The contour levels are the same as those of figure \ref{H2map1}; 0.2, 0.4, 0.8, 1.6, 3.2, 6.4, and 12.8 $M_\odot$ pc$^{-2}$. \label{H2map2}}
\end{figure*}

Figure \ref{H2map2} shows a face-on map of the molecular gas when the conversion factor is calculated with equation (\ref{conv}). In this map, the molecular gas distribution in the outer region is slightly enhanced. 
However, there is no large difference between these maps. 
Hence, our discussions below about the structure of the molecular gas distribution are based on the H$_2$ map obtained adopting a constant conversion factor for simplicity.  

\vskip 3mm
\centerline{--- Figure \ref{H2map1} ---}
\vskip 3mm

\vskip 3mm
\centerline{--- Figure \ref{H2map2} ---}
\vskip 3mm

\vskip 3mm
\centerline{--- Figure \ref{H2arms} ---}
\vskip 3mm

\subsection{Vertical Sliced Maps}
Figures \ref{vertical1} and \ref{vertical2} present vertical cross sections of the resultant cube sliced with planes which are parallel to the $z$-axis. We made 6 slices at the azimuthal angles in each figure. Panels of azimuth angles of $\theta = 40\arcdeg$, $60\arcdeg$, $80\arcdeg$, $100\arcdeg$, $120\arcdeg$, and $140\arcdeg$ in figure \ref{vertical1} are in exactly the opposite directions: $\theta = 220\arcdeg$, $240\arcdeg$, $260\arcdeg$, $280\arcdeg$, $300\arcdeg$, and $320\arcdeg$ in figure \ref{vertical2}. 
\vskip 3mm
\centerline{--- Figure \ref{vertical1} ---}
\vskip 3mm
\vskip 3mm
\centerline{--- Figure \ref{vertical2} ---}
\vskip 3mm

\section{Structure of the Molecular Disk}
\subsection{Radial Distribution}
\begin{figure*}
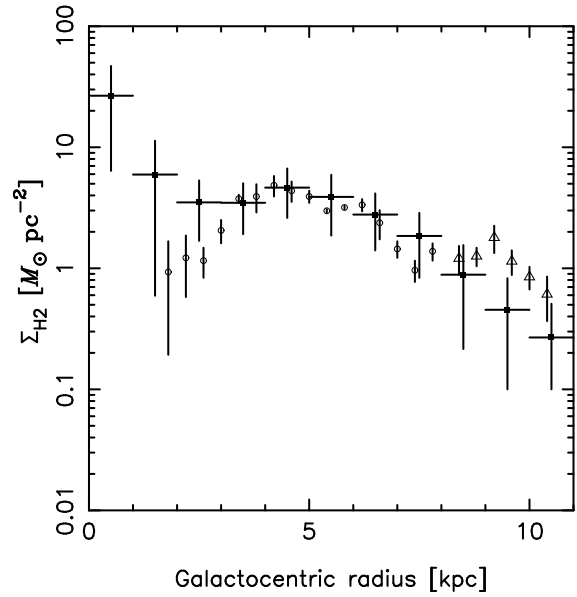

   \begin{center}
    \FigureFile(75mm,150mm){hnakanis-f10.ps}
  \end{center}
  \hspace{1.0cm} 
\caption{Radial profiles of the H$_2$ gas surface density. The vertical axis is a logarithmic scale. Filled squares, open circles, and open triangles denote the data from this study, \citet{bro88} and \citet{gra87}, respectively. \label{radpro-surf-dens}}
\end{figure*}

Figure \ref{radpro-surf-dens} shows a radial distribution of the surface density averaged azimuthally over the 1 kpc-width radius when the constant conversion factor was adopted. {For a comparison, figure \ref{radpro-surf-dens} also shows radial distributions calculated by \citet{bro88} and \citet{gra87}. The radius is reduced by factor of $8/10$ because of difference in the adopted Galactic constant. Moreover, we adopted the conversion factor of $1.8 \times 10^{20}$ although  \citet{bro88} and \citet{gra87} used $X$ of $2.8 \times 10^{20}$. Hence, we reduced the surface density by factor of $1.8/2.8$. The data from \citet{bro88} is a mean value of the northern and southern data. Filled squares, open circles, and open triangles denote the data from this study, \citet{bro88} and \citet{gra87}, respectively. } The surface density decreases from 26.7 $M_\odot$ pc$^{-2}$ to 0.3 $M_\odot$ pc$^{-2}$ with the Galactocentric distance. {Our result is almost consistent with \citet{bro88} and \citet{gra87}.} The plotted values are given in table \ref{radpro-tbl}. 

Radial distributions of the molecular gas in spiral galaxies are often expressed with exponential functions \citep{you95}. 
However, this figure shows that the radial profile is not expressed by an exponential function. 
The surface density is quite high at the Galactic center and near the 4 kpc radius, which is in response to the central condensation and Four-kpc ring found in the face-on map. 
The Galaxy is known to have a bar in the central region (e.g., Nakada et al. 1991). 
These double peaks are thought to relate to the barred structure  (e.g., Nakai 1992).

\vskip 3mm
\centerline{--- Table \ref{radpro-tbl} ---}
\vskip 3mm

\vskip 3mm
\centerline{--- Figure \ref{radpro-surf-dens} ---}
\vskip 3mm

\begin{figure*}
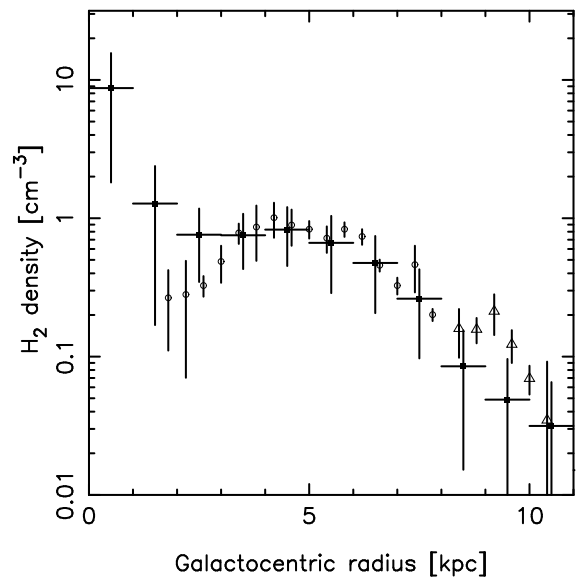

 
   \begin{center}
    \FigureFile(75mm,150mm){hnakanis-f11.ps}
  \end{center}
  \hspace{1.0cm} 
\caption{Radial profiles of the H$_2$ gas density at the midplane. The vertical axis is a logarithmic scale. Filled squares, open circles, and open triangles denote the data from this study, \citet{bro88} and \citet{gra87}, respectively. \label{radpro-midpl-dens}}
\end{figure*}

We also show the radial distribution of the molecular gas density at the midplane in figure \ref{radpro-midpl-dens}. {Radial distributions derived by \citet{bro88} and \citet{gra87} are plotted for a comparison. The radius is reduced by factor of $8/10$, and the density is multiplied by $(1.8/2.8) \times (10/8)^2$ because of the difference in the adopted Galactic constant and the conversion factor. Filled squares, open circles, and open triangles denote the data from this study, \citet{bro88} and \citet{gra87}, respectively.}  The radial distribution is similar to that of the surface density.  The density decreases from 8.73 cm$^{-3}$ to 0.03 cm$^{-3}$ with the Galactocentric distance. {Our result is almost consistent with \citet{bro88} and \citet{gra87}}. The radial profile of this value is not expressed by an exponential function, either. The plotted values are given in table \ref{radpro-tbl}. 

\vskip 3mm
\centerline{--- Figure \ref{radpro-midpl-dens} ---}
\vskip 3mm

\subsection{Scale-height}
Figure \ref{sh-co} shows that the FWHM of the molecular gas disk slightly increases from $48$ pc to $160$ pc with the Galactocentric radius in the radius range of 0--11 kpc. The scale-height is smaller than that of the H {\sc i} disk. 
The plotted values are given in table \ref{radpro-tbl}.

\subsection{Midplane Displacement}
Figures \ref{vertical1} and \ref{vertical2} show that the midplane of the inner disk is slightly displaced from the Galactic plane, which is known as a midplane displacement, a tilted disk, or warped disk. The warping of the outer disk was clearly delineated by \citet{may97}.

\begin{figure*}
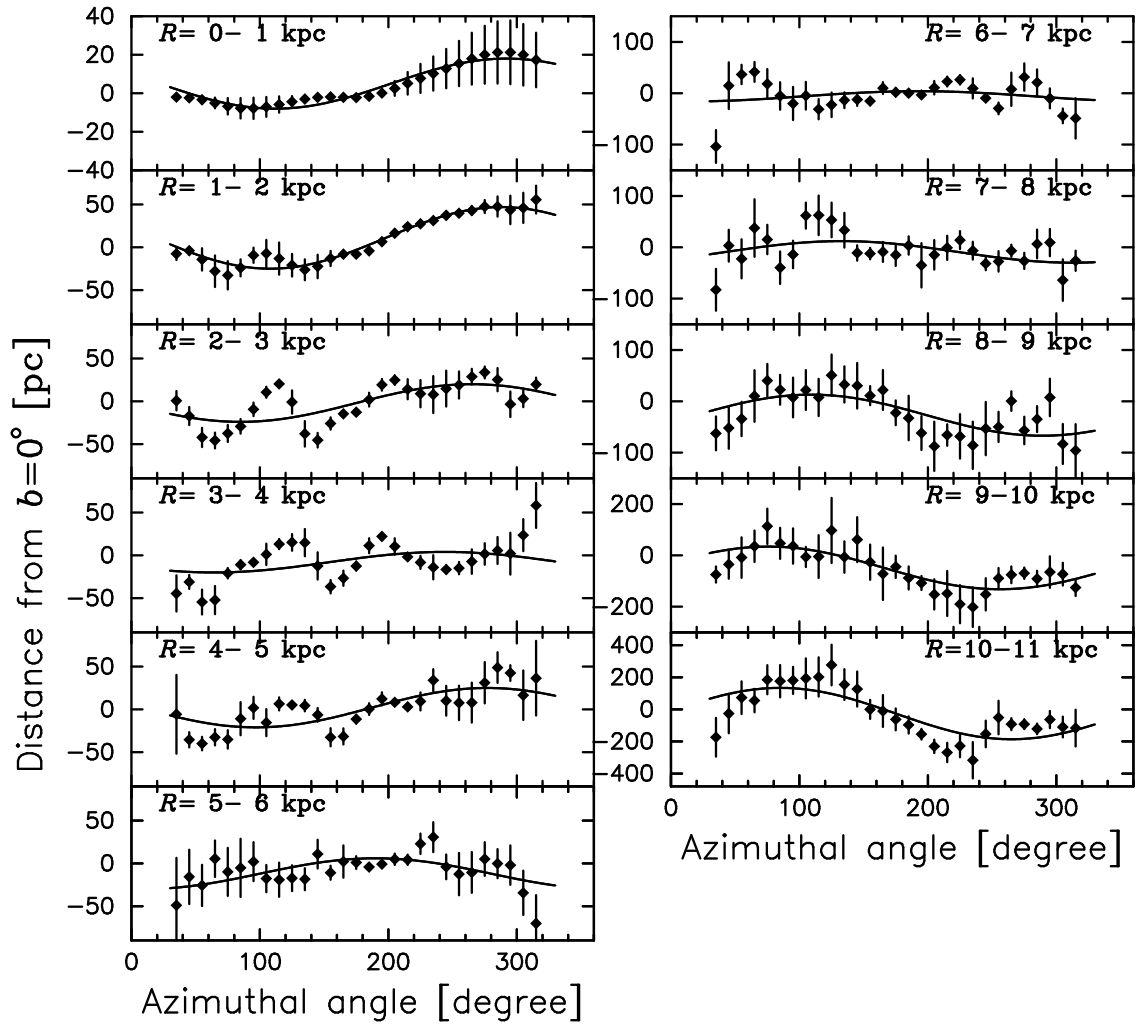

  \begin{center}
    \FigureFile(150mm,250mm){hnakanis-f12.ps}
   \end{center}
\caption{Midplane displacements for individual radii. Horizontal and Vertical axes denote the azimuthal angle around the Galactic center and the distance from the plane of $b=0\arcdeg$. \label{midplane}}
\end{figure*}

Figure \ref{midplane} shows the midplane displacement versus the azimuthal angle $\theta$ relative to the Galactic center for individual radii. The vertical axis denotes the distance between the midplane of the molecular disk and the plane of $b=0\arcdeg$. 
{We calculated the midplane displacement, $z_m$, by averaging $z$ with the weight of density, $n(z)$,
\begin{equation}
z_m = {\int z n(z) dz \over \int n(z) dz}. 
\end{equation} 
}

\begin{figure*}
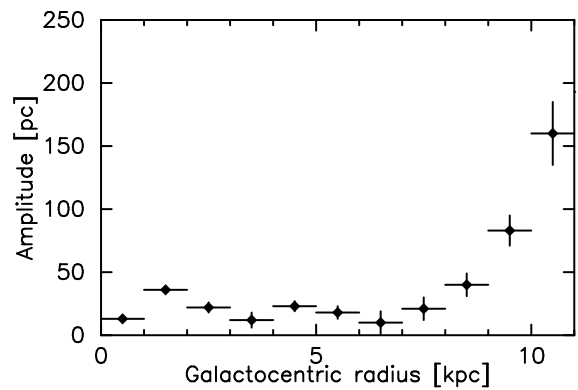

  \begin{center}
    \FigureFile(75mm,75mm){hnakanis-f13.ps}
   \end{center}
\caption{The amplitude of the midplane displacement versus the Galactocentric distance. \label{warpamp}}
\end{figure*}

The midplane displacement can be expressed with a sinusoidal function. Hence, the molecular gas disk can be considered as a collection of tilted rings. We fitted the following sinusoidal function to the obtained data, 
\begin{equation}
z_m = z_0 + A \sin(\theta-\theta_0). 
\end{equation}
The fitting algorithm used in this paper is the nonlinear least-squares Marquardt-Levenberg algorithm used in a software gnuplot. The obtained parameters are given in table \ref{midplane-parm}. 
Figure \ref{warpamp} shows that amplitude of the midplane displacement is nearly constant inside a radius of 10 kpc. However, the amplitude steeply increases beyond the radius of 10 kpc, where the H {\sc i} warp is remarkable \citep{nak03}. 

\begin{figure*}
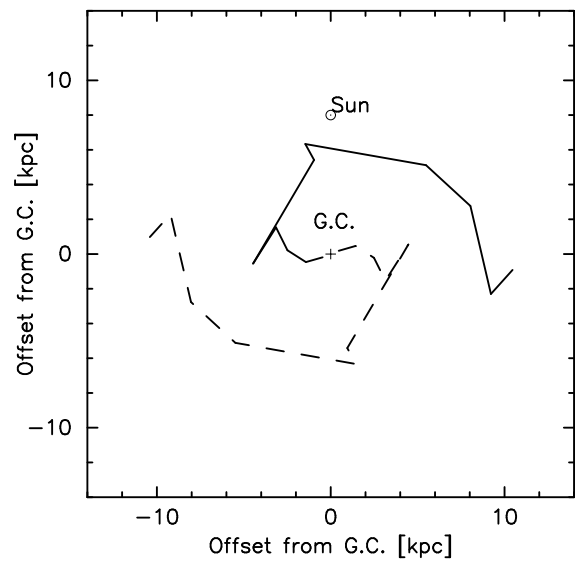

  \begin{center}
    \FigureFile(75mm,75mm){hnakanis-f14.ps}
   \end{center}
\caption{Loci where the midplane displacements reach the maximum in the face-on view. The thick and dashed lines denote the loci where the midplane is displaced upward and downward, respectively. \label{md-maxpoints}}
\end{figure*}

Figure \ref{md-maxpoints} shows loci of the maximum points of individual tilted rings. The solid and dashed lines denote the loci where the midplane is displaced upward and downward, respectively. The loci form leading spiral arms, which is consistent with the rules of the warp summarized by \citet{bri90}, and which is interpreted as follows: 
The midplane displacement can be considered as a wave, called a bending wave \citep{bin87}. Defining $\Omega$ and $\nu$ as the rotational and vertical frequencies, respectively, 
the pattern speed of the bending wave is expressed as $\Omega + \nu/m$ or $\Omega - \nu/m$ ($m$: integer). 
Since the first one denotes a fast propagating wave and a trailing arm, it disappears rapidly as a result of the winding process. On the other hand, the leading spiral component is long-lived. 

\vskip 3mm
\centerline{--- Table \ref{midplane-parm} ---}
\vskip 3mm

\subsection{Central Concentration}
A highly concentrated molecular gas, which is called the central molecular zone (CMZ; Morris \& Serabyn 1996), is {seen} at the Galactic center. The gaseous column density reaches $\gtrsim 25 M_\odot$ pc$^{-2}$ at the center when we adopt a conversion factor of $1.8\times 10^{20}$ cm$^{-2}$ K$^{-1}$ km s$^{-1}$.  Though it is plausible that abundant molecular gas is distributed near the Galactic center, the shape shown in this map is suspicious, because the non-circular motion is noticeable in the central region due to the central Galactic bar. {\citet{saw04} created a face-on view of the CMZ using the data from the CO ($J=$1--0) survey of the Galactic central region \citet{bit97}. Because they derived it using without using the kinematic distance, their map is more reliable. The molecular gas in the CMZ is elongated in the line-of-sight direction. This feature is thought to be associated with a central galactic bar. }

\subsection{Spiral Arms}
\subsubsection{Scutum-Crux Arm}
Positions of the edges of the Scutum-Crux arm in the first and fourth quadrants are $l=33\arcdeg$ and $l=305\arcdeg$ \citep{geo76}.
In the direction of $l=305\arcdeg$, an arm-like structure can be detected around $(R, \theta) \sim (7 \mbox{ kpc}, 250\arcdeg)$, which corresponds to a part of the Scutum-Crux. 
In figure \ref{H2arms}, a schematic tracer of the Scutum-Crux arm is presented by a logarithmic spiral arm starting at $(R, \theta)=(3.0 \mbox{ kpc}, 30\arcdeg)$ with a pitch angle of $12\arcdeg$. The parameters of these logarithmic spiral arms used in figure \ref{H2arms} are given in table \ref{arm-parm}. 

\vskip 3mm
\centerline{--- Table \ref{arm-parm} ---}
\vskip 3mm

\subsubsection{Sagittarius-Carina Arm}
The Sagittarius-Carina arm is a famous arm located near the Sun and is known to arc in the first and fourth quadrants (e.g., Georgelin \& Georgelin, 1976; Cohen et al. 1985; Dame et al. 1986; Grabelsky et al. 1987; Grabelsky et al. 1988; Clemens, 1988; Solomon \& Rivolo, 1989). According to \citet{geo76}, apparent edges of this arm are $l=50\arcdeg$ and $l=283\arcdeg$. 
This arm can be traced by a logarithmic spiral arm with a pitch angle of $11\arcdeg$. 
In figure \ref{H2arms}, a schematic tracer of the Sagittarius-Carina arm is presented by a logarithmic spiral arm starting at $(R, \theta)=(2.0 \mbox{ kpc}, 180\arcdeg)$ with a pitch angle of $11\arcdeg$. 

\subsubsection{Perseus Arm}
A clear spiral structure can be found to arc from $(R,\theta) \sim (7 \mbox{ kpc}, 90\arcdeg)$ to $(R,\theta)\sim (11 \mbox{ kpc}, 170\arcdeg)$. This spiral arm is the so-called Perseus arm. This arm goes by the Sun outside the Solar orbit. The shape of this arm in our map is consistent with the result of \citet{cle88}. A cross section of this arm is found at $R=10-11$ kpc in the top panel of figure \ref{vertical1}. This arm can be traced more clearly in $l-v$ diagram of \citet{dam01} and is detected from $l\sim 55\arcdeg$ in the negative velocity to $ l\sim 270\arcdeg$ in the positive velocity.    

This arm traced in the molecular gas map is consistent with that detected in the H {\sc i} map (Paper I). It is more prominent in the H {\sc i} map. The pitch angle of this arm was estimated to be $15\arcdeg$.  
In figure \ref{H2arms}, a schematic tracer of the Perseus arm is presented by a logarithmic spiral arm starting at $(R, \theta)=(5.7 \mbox{ kpc}, 30\arcdeg)$ with a pitch angle of $15\arcdeg$.

\subsubsection{Norma-Outer Arm}
At larger radius than the Perseus arm, the Outer arm can be traced from $(R,\theta) \sim (9 \mbox{ kpc}, 50\arcdeg)$ to $(R,\theta)\sim (11 \mbox{ kpc}, 120\arcdeg)$. The Outer arm can be seen also in the longitude-velocity diagram of the CO survey \citep{dam01}. This arm appears in the Galactic longitude range of $ 30\lesssim l \lesssim 90\arcdeg$. The Outer arm is also detected in the H {\sc i} map similarly to the Perseus arm. This arm traced in the molecular gas map is also consistent with that detected in the H {\sc i} map, and is more prominent in the H {\sc i} map. This arm can be traced with a logarithmic spiral arm with the pitch angle of $15\arcdeg$.

The Outer arm can be found in the vertical sliced map (figure \ref{vertical2}). The Outer arm is located at the radius of $R\sim 9 - 12$ kpc in the $\theta = 80\arcdeg$, $120\arcdeg$, and $140\arcdeg$ panels of figure \ref{vertical2}. Particularly in the $\theta = 120\arcdeg$ and $140\arcdeg$ panels, this arm is displaced from the Galactic plane due to warping.

The Norma arm is known to exist in the fourth quadrant (e.g., Georgelin \& Georgelin, 1976). \citet{geo76} stated that the apparent edge of this arm is $l=327\arcdeg$. In the fourth quadrant of our H$_2$ gas map, there is an elongated lump with a high H$_2$ column density in the direction of $l=327\arcdeg$, which corresponds to the Norma arm. Though \citet{geo76} mentioned that the Norma arm is symmetrical to the Sagittarius-Carina arm, the Norma arm seems symmetrical to the Scutum-Crux arm rather than the Sagittarius-Carina arm. 

These two spiral arms seem to be the same arm, assuming the pitch angles of these arms are $15\arcdeg$. Hence, we call these arms the Norma-Outer arm. {There are other possibilities of connections of (i) the Norma and Perseus arms and (ii) the Scutum-Crux and Outer arms. However, the radius of the Perseus is expected to be too small to be connected to the Norma arm in the fourth quadrant. Similarly, the radius of the Scutum-Crux arm is too large to be connected to the Outer arm in the first quadrant. Therefore, the connection of the Norma and Outer arms seems most natural. }

In figure \ref{H2arms}, a schematic tracer of the Norma-Outer arm is presented by a logarithmic spiral arm starting at $(R, \theta)=(4.0 \mbox{ kpc}, 240\arcdeg)$ with a pitch angle of $15\arcdeg$.

\subsubsection{Comparison with arms identified from HII region data}
\begin{figure*}
  \begin{center}
    \FigureFile(150mm,150mm){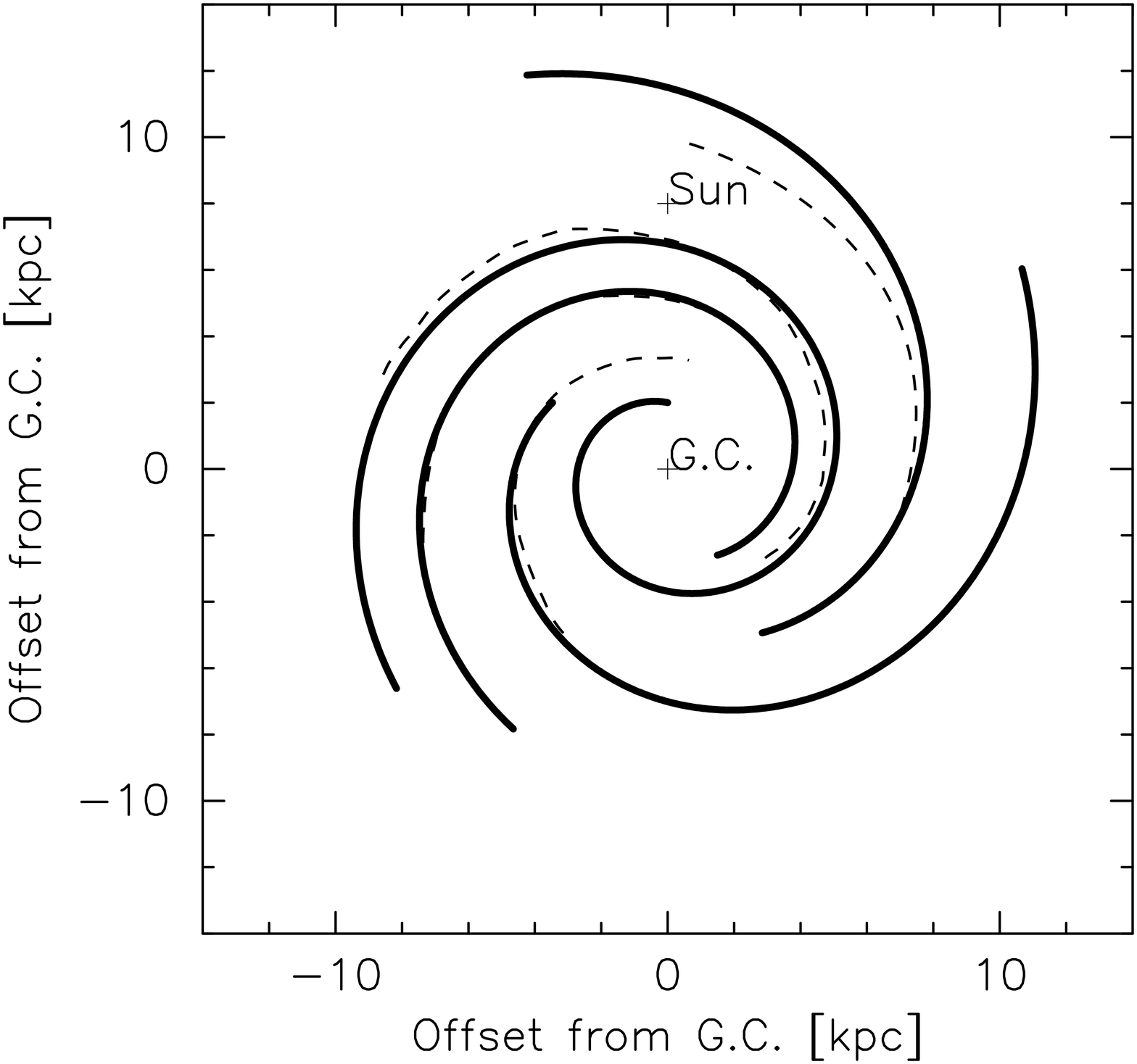}
   \end{center}
\caption{Comparison of our result and spiral structure presented by \citet{geo76}. Thicker lines denote spiral arms which we traced. Thinner dashed lines denote spiral arms which \citet{geo76} traced. \label{comp-geo76}}
\end{figure*}

{In order to evaluate the validity of the spiral arm that we trace, we compare our result with \citet{geo76}. We overlay the spiral arms that \citet{geo76} traced on our tracing spiral arms in figure \ref{comp-geo76}. The sizes of spiral arms are changed so that the distance between the Galactic center and the Sun is the same. The Sagittarius-Carina, the Scutum-Crux, and the Norma arms that we trace are coincident with those traced by \citet{geo76}, although the radius of the Perseus arm is slightly different for $l \gtrsim 90\arcdeg$ in the cases of our result and \citet{geo76}.}

\subsection{Ring-like structure}
In the first quadrant ($0\arcdeg \le l \le 90\arcdeg$) a ring-like feature is found at $R\sim 4$ kpc, which corresponds to the ring-like structure found by \citet{cle88}. 
The difference in the scale of this structure is due to the adopted $R_0$. \citet{cle88} adopted $R_0 = 10$ kpc, while we adopted  $R_0 = 8$ kpc. Hence, the scale of our map is 4/5 times as large as the map of \citet{cle88}. 
We call this structure the Four-kpc Ring instead of the Five-kpc ring in this paper.

The edge of the Four-kpc ring corresponds to $l=30\arcdeg - 40\arcdeg$. This coincides with the edge of the Scutum-Crux arm. 
In addition, the Sagittarius-Carina arm seems to be merged into the Four-kpc Ring in the azimuthal range of $\theta\lesssim 120\arcdeg$. The Four-kpc ring may therefore be a complex of two spiral arms and remarkably enhanced as if it was a ring.

{\citet{bro88} pointed out that the radial distribution of molecular gas is flatter in the south hemisphere than in the north, and that the density peak at $0.5 R_0$  in the north broadens in the south into a plateau. This statement is consistent with our conclusion that two spiral arms in the south are merged in the north. }

\section{Possible Errors in our Molecular Gas Map}
Since the molecular gas map we obtained in this study is based on several assumptions, we must note the systematic errors due to deviations from these assumptions. 

First, we assumed that the Milky Way galaxy rotates absolutely circularly around the Galactic center. However, the gaseous motion deviates from circular motion due to the density wave or the central Galactic bar, believed to exist at the Galactic center. The isovelocity contours are typically distorted in 1 kpc according to observations of external galaxies. Uncertainties concerning the location of the gas are about 1 kpc.

Second, we used the model expressed by equation (\ref{spi_eq}) to divide the mixture emission from near and far points. Though the resultant map can change due to the adopted function, the global feature is almost the same for any function as shown in Paper I. 

{Third, fitting method does not necessarily work. Typical residual is less than 20 -- 30 \% of midplane density (figure \ref{residualmap}). However, residual becomes much larger near the Sun, which implies that the fitting method does not work well in this region. }

Fourth, conversion factor also affects the resultant gaseous map. We tried to make 3-D maps using a constant conversion factor and a varying conversion factor with the Galactocentric distance. Outer structure is enhanced when we use the varying conversion factor. 

\section{SUMMARY}
We constructed a three-dimensional molecular gas map of the Milky Way Galaxy using the new {compilation data set of} $^{12}$CO($J=$1--0) survey {obtained with 1.2-m telescopes} and the latest rotation curve, based on the kinematic-distance. We made molecular gas maps by adopting two types of conversion factor; one is constant, and the other varies with the Galactocentric distance. Global features are the same for both though the contrast is slightly different. 
 
{We present a new 3-D map of the molecular gas in this paper.} The resultant maps {confirm} the following features of the Galactic molecular disk. 
\begin{enumerate}
\item The radial distribution of the molecular gas surface density has double peaks at the center and at the radius of $0.5R_0$, which is a feature similar to barred spiral galaxies. 
\item The FWHM of the molecular disk slightly increases from 48 pc to 160 pc with the Galactocentric distance within the radius range of 0--11 kpc.
\item There is a midplane displacement, whose amplitude is nearly constant inside the radius of 10 kpc, increasing beyond this radius. The ridges of midplane displacement form leading spiral arms. 
\item Molecular gas arms corresponding to the Outer, the Perseus, the Sagittarius-Carina, the Scutum-Crux, and the Norma arms could be traced as logarithmic spiral arms with the pitch angle of $11\arcdeg - 15\arcdeg$. Supposing that the pitch angles of the spiral arms are within this range, the Norma can be identified as being the same as the Outer arm. 
\end{enumerate}

\bigskip

We are grateful to Dr. Thomas Dame et al. for providing us with their large CO survey data.

\begin{figure*}
  \begin{center}
    \FigureFile(150mm,150mm){hnakanis-f8.ps}
   \end{center}
\caption{Vertical cross--section maps. Each map shows the H$_2$ volume densities in a sheet through the Galactic center perpendicular to the Galactic plane. First plate: Individual panel shows sections through $\theta=220\arcdeg$, $\theta=240\arcdeg$, $\theta=260\arcdeg$, $\theta=280\arcdeg$, $\theta=300\arcdeg$, and  $\theta=320\arcdeg$. 
The contours are drawn at levels of 0.05, 0.1, 0.2, 0.4, 0.8, 1.6, and 3.2 H$_2$ cm$^{-3}$.  \label{vertical1}} 
\end{figure*}

\begin{figure*}
  \begin{center}
    \FigureFile(150mm,150mm){hnakanis-f9.ps}
   \end{center}

\caption{Vertical cross--section maps. Each map shows the H$_2$ volume densities in a sheet through the Galactic center perpendicular to the Galactic plane. Individual panel shows sections through $\theta=40\arcdeg$, $\theta=60\arcdeg$, $\theta=80\arcdeg$, $\theta=100\arcdeg$, $\theta=120\arcdeg$, and  $\theta=140\arcdeg$. The contours are drawn at levels of 0.05, 0.1, 0.2, 0.4, 0.8, 1.6, and 3.2 H$_2$ cm$^{-3}$. \label{vertical2}} 
\end{figure*}


\begin{figure*}
  \begin{center}
   \rotatebox{-90}{\FigureFile(150mm,150mm){hnakanis-f7.ps}}
   \end{center}
\caption{Schematic tracers of H$_2$ gas arms superimposed on the obtained H$_2$ column density map; No.1: Norma-Outer arm; No.2: Perseus arm; No.3: Sagittarius-Carina arm; No.4; Scutum-Crux arm. The contour levels are the same as those of figure \ref{H2map1}; 0.2, 0.4, 0.8, 1.6, 3.2, 6.4, and 12.8 $M_\odot$ pc$^{-2}$. \label{H2arms}}
\end{figure*}

\begin{table*}
\begin{center}
\caption{Radial variations of parameters of the molecular disk  \label{radpro-tbl}}
$\begin{array}{D{-}{-}{-1}D{+}{\pm}{-1}D{+}{\pm}{-1}D{+}{\pm}{-1}}
\hline\hline
\multicolumn{1}{c}{\rm Radius} & \multicolumn{1}{c}{\mbox{Surface Density}} & \multicolumn{1}{c}{\mbox{Midplane Density}} & \multicolumn{1}{c}{\rm FWHM} \\
\multicolumn{1}{c}{\rm (kpc)} & \multicolumn{1}{c}{(M_\odot {\rm pc}^{-2})} & \multicolumn{1}{c}{({\rm cm}^{-3})} & \multicolumn{1}{c}{({\rm pc})} \\
\hline
 0 - 1 & 26.7+ 20.3 & 8.73+ 6.91  & 48+20 \\
 1 - 2 &  6.0+  5.4 & 1.28+ 1.11  & 56+18 \\
 2 - 3 &  3.5+  1.8 & 0.76+ 0.42 & 68+14  \\
 3 - 4 &  3.5+  1.6 & 0.75+ 0.32  & 88+24 \\
 4 - 5 &  4.6+  2.0 & 0.83+ 0.38  & 78+12 \\
 5 - 6 &  3.9+  2.0 & 0.66+ 0.38 & 102+10  \\
 6 - 7 &  2.8+  1.4 & 0.48+ 0.27 & 84+22  \\
 7 - 8 &  1.9+  1.0 & 0.26+ 0.17  & 90+30 \\
 8 - 9 &  0.9+  0.7 & 0.09+ 0.07  & 186+20 \\
 9 -10 &  0.5+  0.4 & 0.05+ 0.05   & 182+20\\
10 -11 &  0.3+  0.2 & 0.03+ 0.03  & 160+20\\
\hline
\end{array}$
\end{center}
\end{table*}

\begin{table*}
\begin{center}
\caption{Fitted parameters of the midplane \label{midplane-parm}}
$\begin{array}{D{-}{-}{-1}D{.}{\pm}{-1}D{.}{\pm}{-1}D{.}{\pm}{-1}}
\hline\hline
\multicolumn{1}{c}{\rm Radius} & \multicolumn{1}{c}{z_0} & \multicolumn{1}{c}{A} & \multicolumn{1}{c}{\theta_0} \\
\multicolumn{1}{c}{\rm (kpc)} & \multicolumn{1}{c}{\rm{(pc)}} & \multicolumn{1}{c}{\rm{(pc)}} & \multicolumn{1}{c}{(\arcdeg)} \\
\hline
0-1 & 5.1 & 13.1 & 292.3 \\
1-2 & 11.1 & 36.2 & 288.4 \\
2-3 & -2.4 & 22.4 & 265.15 \\
3-4 & -8.5 & 12.6 & 244.32 \\
4-5 & 2.3 & 23.4 & 277.13 \\
5-6 & -12.3 & 18.5 & 190.13 \\
6-7 & -6.6 & 10.9 & 193.43 \\
7-8 & -9.6 & 21.9 & 133.23 \\
8-9 & -27.7 & 40.9 & 109.15 \\
9-10 & -49.10 & 83.12 & 76.11 \\
10-11 & -26.30 & 160.25 & 85.11 \\
\hline
\end{array}$
\end{center}
\end{table*}

\begin{table*}
\begin{center}
\caption{Parameters of schematic spiral arms \label{arm-parm}}
$\begin{array}{lD{.}{.}{-1}D{.}{.}{-1}D{.}{.}{-1}D{.}{.}{-1}D{.}{.}{-1}}
\hline\hline
\multicolumn{1}{c}{\rm Name} & \multicolumn{1}{c}{\rm Pitch Angle} &\multicolumn{2}{c}{\mbox{Beginning Point}} & \multicolumn{2}{c}{\mbox{Ending Point}}\\
\multicolumn{1}{c}{ } & \multicolumn{1}{c}{(\arcdeg)} &  \multicolumn{1}{c}{\rm{(kpc)}} & \multicolumn{1}{c}{(\arcdeg)} & \multicolumn{1}{c}{\rm{(kpc)}} & \multicolumn{1}{c}{(\arcdeg)}\\
\hline
\mbox{Scutum-Crux}        & 12 & 3.0 &  30 &  9.1 & 330 \\
\mbox{Sagittarius-Carina} & 11 & 2.0 & 180 & 10.5 & 310 \\
\mbox{Perseus}            & 15 & 5.7 &  30 & 12.6 & 200 \\
\mbox{Norma-Outer}        & 15 & 4.0 & 240 & 12.3 & 120 \\
\hline
\end{array}$
\end{center}
\end{table*}

\end{document}